# Design and Implementation of a Low-Cost Secure Vehicle Tracking System

Ibraheem Kasim Ibraheem
Electrical Engineering Department
College of Engineering, Baghdad University
P.O.B.: 47273, 10001, Baghdad, Iraq
ibraheemki@coeng.uobaghdad.edu.iq

Salam Wisam Hadi
Electrical Engineering Department
College of Engineering, Baghdad University
10001, Baghdad, Iraq
salam_skull007@yahoo.com

*Abstract*— **In this paper, we used XBee wireless technology due to the privileges that it provides in terms of low cost and a high level of security which gives a more reliable information transfer, penetration avoidance, and unauthorized access, without any cost in sending and receiving this information. The aim of this work is to syndicate the XBee wireless technology and global positioning system (GPS) for a low-cost real-time vehicle tracking system and displaying the result on Google earth. The overall system involved two main modules, the displaying module (monitoring station) and the following module (vehicle unit). The following module consists of microcontroller (Arduino) platform, XBee, and GPS for navigation purpose. The GPS delivers real-time data about the location of the vehicle and directs the coordinate to the XBee via the Arduino platform. The later is incorporated as a connecting buffer between the XBee transmission unit and the GPS receiver. Receiving the location data of the tracked vehicle and displaying them on Google earth is the responsibility of the monitoring station. The designed system has been tested practically in both crowded and open area environments, the overall system works well and displayed the vehicle coordinates nevertheless the existence of noise and interference in the vehicle area and regardless the obstacles like buildings.**

## I. Introduction (Heading 1)

Modern vehicle tracking systems (VTS) commonly use Global Positioning System GPS or Global Navigation Satellite System GLONASS technology for locating the vehicle or global navigation satellite system (Galileo), in any case, different kinds of automatic vehicle locations technologies can likewise be utilized. Vehicle data can be seen on electronic maps through the Web or particular S/W. Urban open travel establishments are an inexorably normal client of VTS, especially in substantial urban communities. Recently, VTS found in many applications, VTS can be used as an anti-theft device, as it helps with the recovery of stolen vehicles by tracking the locations of vehicle movements. Having a VTS can also reduce insurance premiums. For businesses, VTS can be used for fleet management, reducing costs, employee tracking, and asset protection [1]. Basically, there are two main types of VTS according to whether location information is transmitted in real time or not, *passive tracking* and *active tracking*. In passive tracking, location information logged in a small memory inside the tracking device in the vehicle. Once the vehicle returns to a predetermined base, data are manually transferred to a central computer. While in Active tracking the data are transmitted in real-time. The tracking device collects the same information but usually transmit these data in near real-time using cellular or satellite networks to a computer or data center for evaluation and displaying. Modern tracking system use combination of both passive and active tracking capabilities [2, 3].

In recent times, many types of research in VTS have been done; these researches enlightened the most important works in this field, few of them are described. The work in [4] proposed and carried out a low cost VTS based on GPRS, GPS, and GIS. The vital H/W and S/W components of the system are an open-source GIS testbed, HTTP protocol, a web application in view of PHP, JavaScript, MySQL with the Google Map embedded, communication server, database server, and a map server. An anti-theft system utilizing an embedded system employed with a GPS and GSM was proposed in [5], The client interacts through this system with vehicles and determines their current locations and status using Google Earth. Application of VTS in anti-theft of vehicles are mentioned in [6, 7, 8]. The main principle of the scheme proposed in [9] was to plan an emergent tracking method, the proposed system was implemented based on a real-time embedded system for this implementation we used ATMEG328P controller which is the heart of the system. The IR sensor is used to sense the data which is processed by the controller and sent the data to emergent tracking system through GPS. Implementation of a vehicle tracking system using smartphone and SMS service was detailed in [10]. On the other hand, XBee is a speciation of a joint of high-level wireless communication protocols based on the wireless Personal Area Network (PAN) standard IEEE 802.15.4. The characteristics that make it so suitable for nowadays applications are, Low-cost since XBee devices do not need a high data rate, Mesh topology, which provides a higher reliability because multiple transmission paths exist, and Low power consumption, as multiple nodes can be asleep until they receive some information. In [11] people introduced a method of integrating a low-cost sensor device (XBee) and Global Positioning System, the proposed system provides real-time tracking application that provides the location of the mobile unit to the fixed control unit as well as display advertisements. While [12] presented the idea of combining (XBee) modules and GPS with multi-mode (Wi-Fi/3G) information and communication technologies.

The key Features of XBee devices are, it can operate globally in the 2.4 GHz frequency, but also in 868 MHz and 915 MHz, Its data rate is 250 kbps at 2.4 GHz, 20 kbps at 868 MHz and 40 kbps at 915 MHz. It operates over 16 channels in 2.4 GHz and over 11 channels in 868 and 915 MHz. There are different ranges of XBee from 300 Ft to 40 Miles, the different distances depending on the type of device, as well as differences and frequency data Rate and antenna and specifications generally. Finally, a XBee network can have a maximum of 255 nodes can its topology can be either in star, mesh, and tree.

Concerning security, XBee provides facilities for carrying out secure communications, protecting establishment and transport of cryptographic keys, cyphering frames and controlling devices. It builds on the basic security framework defined in IEEE 802.15.4. XBee uses 128-bit keys to implement its security mechanisms. The security architecture is distributed among different protocol layers, The MAC sublayer, The network layer, and The application layer.

The main key points of this work are summarized in the following:

1. Studying and analyzing the tracking system of the vehicle using GPS.
2. Designing a secure mechanism for the vehicle tracking system using XBee transmission and receiving principles.
3. Implementation the designed secure vehicle tracking system practically using Arduino microcontroller.
4. Testing the designed system on a real vehicle and mapping the obtained Coordinates on Google Earth.

The paper structure is as follows: Section II presents the conventional vehicle tracking system(VTS). The proposed configuration of the VTS using XBee technology is introduced in Section III. The simulation results are included in Section IV. The paper is concluded in Section V.

## II. CONVENTIONAL VTS BASED ON GPS

In recent times, numerous kinds of research have been accomplished a fused GPS-GSM location finding of VTS. A coordinated GPS-GSM Installed system have been utilized to track the present position of the vehicle by means of "Google Earth". The GPS receiver (MediaTek MT3329), microcontroller and GSM modem (SIM 900D) were attached in the vehicle. The GPS receiver has been utilized to get the signal from the satellite. The microcontroller has been employed to examine the specific engine considerations and transmit wanted information to the server through the GSM modem. The second modem which is associated with the PC will get the SMS that incorporates the GPS coordinate and engine parameters. The Visual Basic program has been adopted to transform the obtained SMS content to numerical frame at that point saved in a "Microsoft Office Excel file". To demonstrate the area of the vehicle and the engine parameters on the map, Microsoft Office Excel file was changed to KML (Keyhole Markup Language) arrangement and Google Earth will translate the KML document. This goal of this system is to deal with an fleet, police autos spreading and vehicle- theft alerts [13], see Fig. 1.

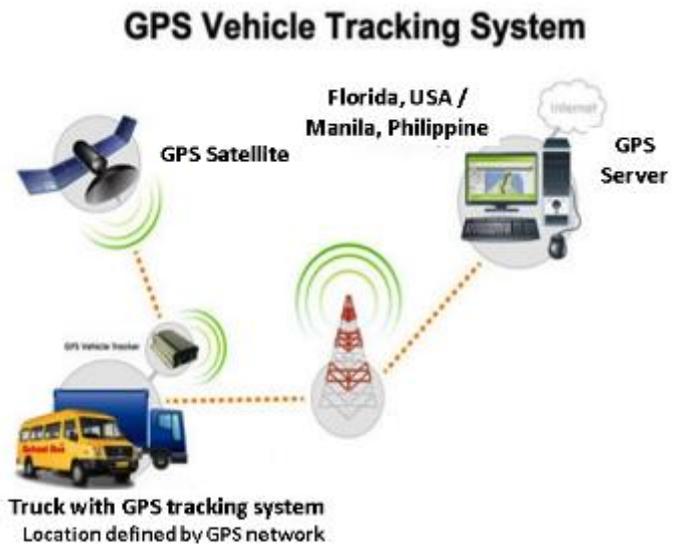

Fig. 1. Conventional Vehicle Tracking System.

## III. DESIGN AND IMPLEMENTATIONS OF SECURE VTS SYSTEM

This section presents the proposed Hardware and Software system design for the secure VTS.

### A. Hardware Components of Secure Vehicle Tracking System

The H/W of the secure Vehicle tracking system is separated into two primary components; the Vehicle-module (following module) and the receiving module (monitoring Station). The following module is in charge of getting the position of the client, while the monitoring Station is for showing the identified position on "Google Earth". The equipment gadgets utilized as a part of this work are, GPS receiver (gy-gps6mv2), Arduino UNO Microcontroller, XBees for Transmitting and receiving the location coordinates, XBee shield, and PC for showing the coordinates on Google earth. Figure 3.3 illustrates the secure VTS components and the interactions between them.

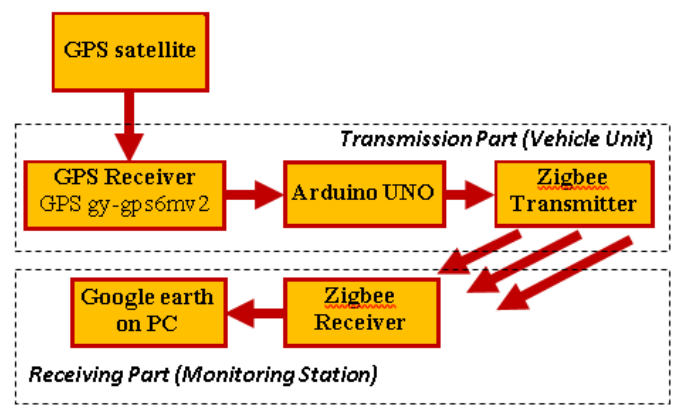

Fig. 2. Secure VTS.

In the transmission side, we first need a connection between GPS receiver and Arduino UNO Microcontroller. The GPS receiver used in this project is (gy-gps6mv2); its Specifications are: Power Supply Range: 3 V to 5 V, Ceramic antenna, EEPROM for saving the configuration data when powered off, Backup battery, LED signal indicator, Antenna Size: 25 x 25 mm, Module Size: 25 x 35 mm, Mounting Hole Diameter: 3 mm, Default Baud Rate: 9600 bps. In the beginning, the GPS device (gy-gps6mv2) gets the information from GPS satellite by National Marine Electronics Association (NMEA) Data, for more information on NMEA, refer to [14]. Then the GPS receiver gets the coordinates and sends them to the UNO Arduino microcontroller. This unit interns excerpts some portion of the "$GPGGA sentences" from the recieved signal and saves it in an array. The "$GPGGA sentence" is one of the NMEA sentences which is ordinarily employed. It has data other than the latitude and longitude, e.g., the time when the signal is received from the GPS, fix quality, number of satellites being followed, elevation, the height of "geoid above WGS 84 ellipsoid" and DGPS reference station ID. Nevertheless, scope and longitude (x, y) are simply required. The connection between different components of the transmission side unit is shown in Fig. 3.

When the system is ON, the microcontroller sends coordinates (*x*, *y*) to the XBee. The second connection in the transmission part is the connection of XBee to Xbee shield and Arduino UNO Microcontroller which is shown in Fig. 3. The

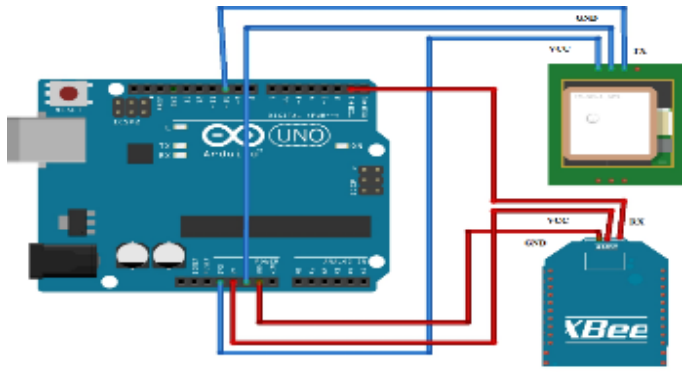

Fig. 3 . Connection of GPS and UNO.

Specifications of XBee modules are: 3.3V , 215mA, 250kbps Max data rate, 60mW output (+18dBm), 1 mile (1500m) range, Built-in antenna, Fully FCC certified, 6 10-bit ADC input pins, 8 digital IO pins, 128-bit encryption, Local or over-air configuration, AT or API command set. Fig. 4 shows the

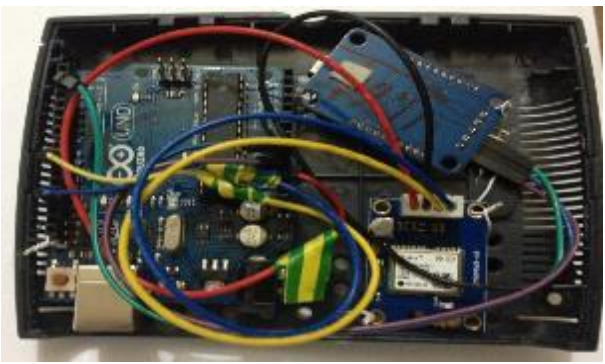

Fig. 4. Connection of XBbee Module to UNO.
schematic of the overall system of the transmission side.

On the monitoring station, we have a XBee module connected to PC through the USB cable, when the XBee module receives the coordinates (x, y) from the (Vehicle-Unit), it will send that coordinates to PC, and shows these (x, y) on Google Earth in a real-time fashion. The connection between the XBee module and the PC is shown in Fig. 5, the setting of Google earth have been modified to display the coordinates (*x*, *y*) in a real-time fashion.

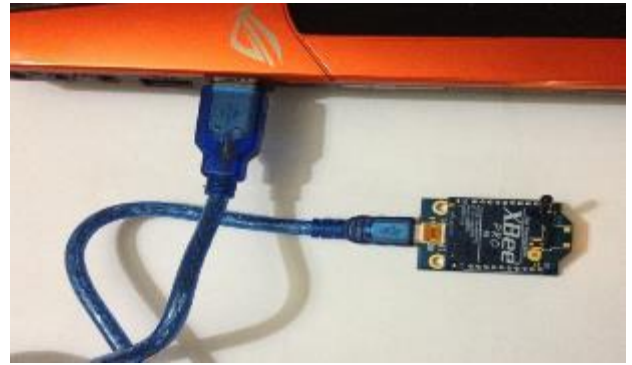

Fig. 5. Connection of Zigbee Module to UNO.

### B. The Software Design of Secure Vehicle Tracking System

The Arduino programming language comes with a software library called "Wiring" from the original Wiring project, which makes many common input/output operations much easier. In this case, we used a GPS library (Tiny GPS-13) and XBee library. To check whether there is a valid communication between the sending and receiving XBee modules, (XCTU) program has been used. Also, it can monitor the communication between them. On the other side, Google Earth settings should be changed to display the coordinates only in a real-time fashion from the tools menu, as follows:

Tools --------> GPS--------> Real-time

Finally, it is best to describe the operation of the entire system by flowchart as shown in Fig. 6.

### IV. SIMULATIONS AND RESULTS

The next stage after finishing the design and implementation of the secure tracking system is the simulation of the complete circuit. It guarantees that the secure VTS is functioning admirably and satisfies the predetermined prerequisites. This section will demonstrate about the outcomes that are acquired from the testing the performance of the overall system, making the necessary simulations using the simulator (SatGen 3) and comparing the results for different distances of the monitoring station. Simulator (SatGen 3), SatGen can be used to create a scenario which replicates very high dynamic conditions, allowing engineers to test their systems work in all environments. This program has been used to simulation for a tracking system that allows showing the results on the simulator, and the obtained curves explain the Coordinates (x, y) position and speed profile per time as shown in Fig. and Fig. 8.

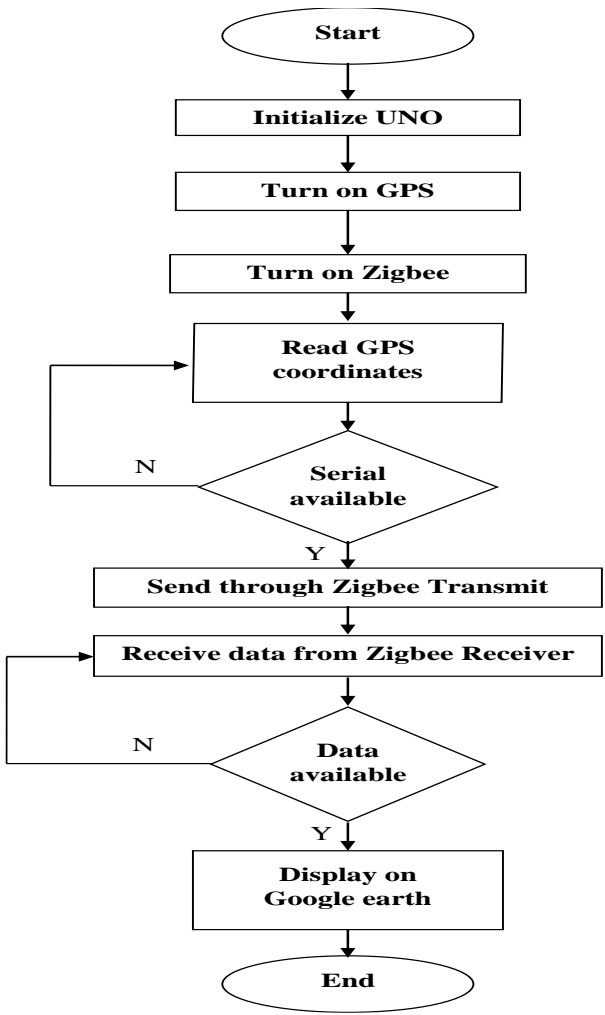

Fig. 6. Flowchart of the complete secure VTS

(SatGen 3) the simulator also gives us the global date and the exact day, Steps of achieving this is as follows are:

1. Make a route from within the program, or insert File (KML).
2. Set time and date.
3. Making simulation to give results.
4. Can change the setting from (User defined) to change speed and height.

Google earth provides options to the user to view road map or satellite view; also zoom in, zoom out, and navigation tools are also available. Route type can be shown in either static or real-time tracking. The overall system can be described as following. The GPS reciever will begin to receive the signal from the atellite after the GPS is turned on. Then it sends the coordinates by XBee modules transmitter. In the monitoring station, XBee modules receiver will receive the coordinates and displays them on Google earth. The program XCTU is software that illustrates whether a stable connection between two XBee modules is achieved or not. Also, it lists the data that are sending by XBee transmitter module to receiving unit its command area. Practically, the overall design system has been experimented in a crowded environment (in terms of buildings and the signals in the air) and in open area to test the tracking system in real-time as shown in Fig. 9 and Fig. 10.

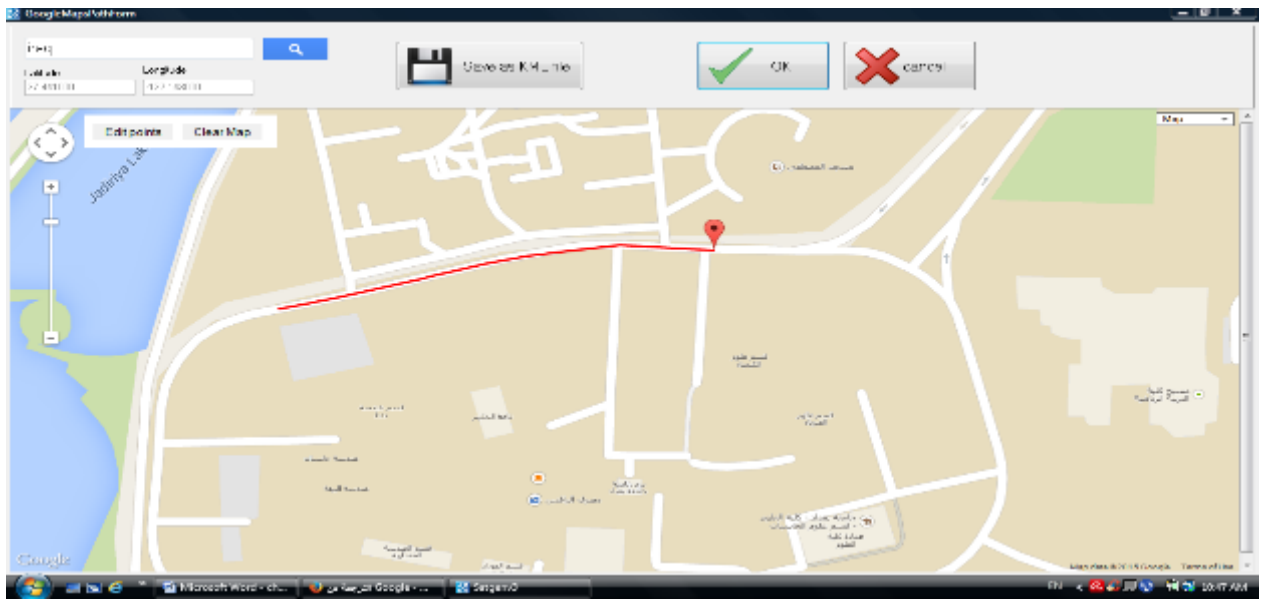

Fig. 7. SatGen 3 Simulator.

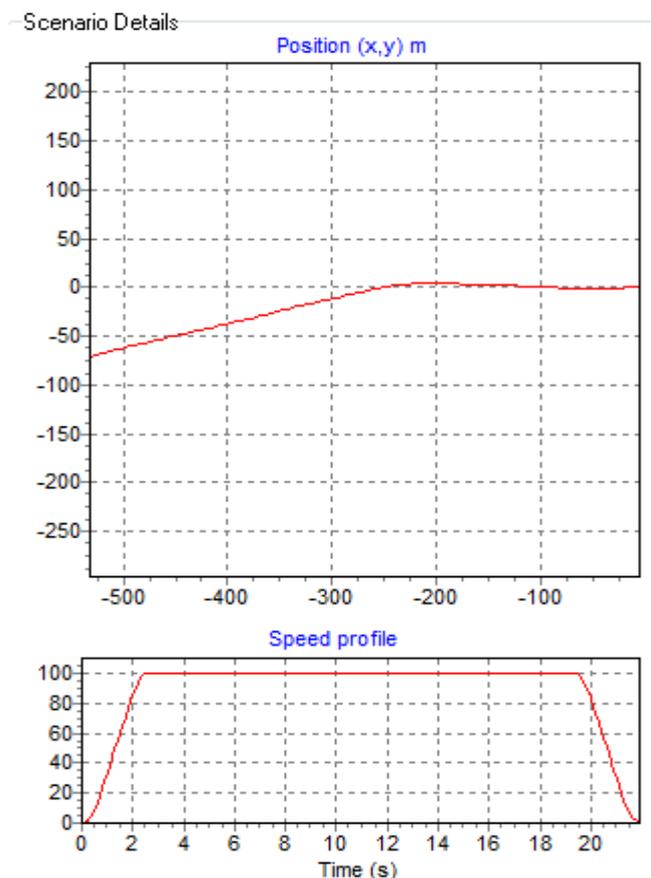

Fig. 8. Position and speed profile of SatGen 3 Simulator.

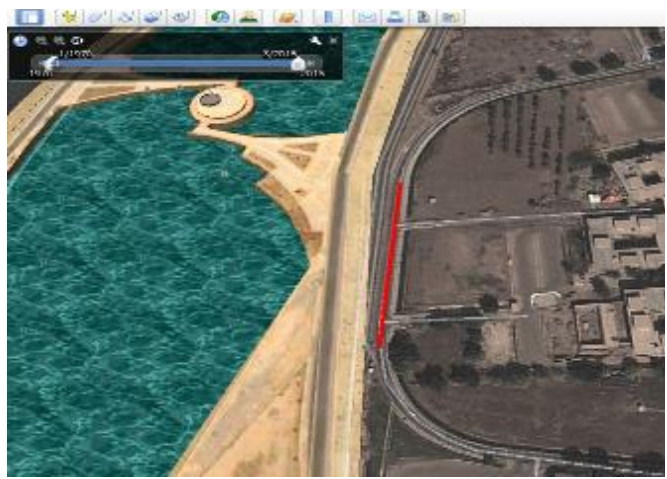

Fig. 10. The tracked route: Test 2.

The following Arduino code is used to show the coordinates on a browser other than the Google earth such as Google maps or GIS. Fig. 11 shows the sample code for this case. However, we just need Coordinates (x, y) in our project to be shown.

```
case 0  :Serial.print("Time in UTC (HhMmSs): ");break;
case 1  :Serial.print("Status (A=OK,V=KO): ");break;
case 2  :Serial.print("Latitude: ");break;
case 3  :Serial.print("Direction (N/S): ");break;
case 4  :Serial.print("Longitude: ");break;
case 5  :Serial.print("Direction (E/W): ");break;
case 6  :Serial.print("Velocity in knots: ");break;
case 7  :Serial.print("Heading in degrees: ");break;
case 8  :Serial.print("Date UTC (DdMmAa): ");break;
case 9  :Serial.print("Magnetic degrees: ");break;
case 10 :Serial.print("(E/W): ");break;
case 11 :Serial.print("Mode: ");break;
case 12 :Serial.print("Checksum: ");break;
```

Fig. 11. Arduino code.

## V. CONCLUSIONS

This work travelled around the basis of Arduino, XBee, and GPS and employed all of these modules to design A secure prototype of VTS.. A low-cost vehicle tracking system was presented. The integration of GPS with XBee provided a continuous and real-time tracking. the main conclusions in this project are, the transmission cost is extremely reduced by using XBee service instead of GSM or GPRS, the overall implementation of the proposed system replaced the traditional GSM/GPRS based tracking system in terms of complexity, the GPS like XBee wireless technology can determine vehicle position tremendously precisely and consistently at locations which have a clear line-of-sight to the satellite (open area) but it does not accomplish well in indoor or in closed environments. Finally, the monitoring station tracked the vehicle well both in dense and open environments and displayed the tracked vehicle coordinates on Google earth without any delay regardless of the noise, interference and the buildings in the vehicle area.

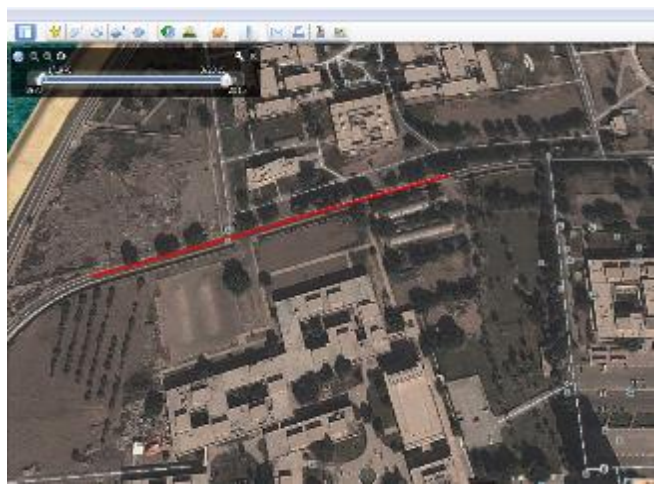

Fig. 9. The tracked route: Test 1.

REFERENCES

[1] http://vehicletracking.expertmarket.co.uk/.

[2] A. M. Ajagbe, S. E. Eluwa, E. E. Duncan, M. K. BinRamliy, C. S. Long, and M. C. Wantrudis, "The Use of Global System of Mobile Communication (GSM) Among University Students in Malaysia," *Int. J. Innov. Manag. Technol.*, vol. 2, no. 6, p. 512, 2011.

[3] http://www.trackingtheworld.com/articles/PassiveVersusActiveGPS.html.

[4] A. Bidgoli and M. Amanifar, "GPS , GPRS , GIS for Tracking System," *Int. J. Comput. Sci. Eng. Technol.*, vol. 1, no. 8, pp. 527–529, 2011.

[5] M. N. Ramadan, M. a Al-khedher, S. Member, and S. a Al-kheder, "Intelligent Anti-Theft and Tracking System for Automobiles," *Int. J. Mach. Learn. Comput. Vol. 2, No. 1,* vol. 2, no. 1, pp. 88–92, 2012.

[6] T. Monawar, S. B. Mahmud and A. Hira, "Anti-theft vehicle tracking and regaining system with automatic police notifying using Haversine formula," *4th International Conference on Advances in Electrical Engineering (ICAEE)*, Dhaka, , pp. 775-7,79 2017.

[7] M. S. Uddin, M. M. Ahmed, J. B. Alam and M. Islam, "Smart Anti-theft vehicle tracking system for Bangladesh based on Internet of Things," *4th International Conference on Advances in Electrical Engineering (ICAEE)*, Dhak, pp. 624-628, 2017.

[8] M. Managuli, A. Deshpande, and S. H. Ayatti, "Emergent vehicle tracking system using IR sensor," *2017 International Conference on Electrical, Electronics, Communication, Computer, and Optimization Techniques (ICEECCOT)*, Mysuru,, pp. 71-74, 2017.

[9] N. Jahan, K. Hossen, and M. K. H. Patwary, "Implementation of a vehicle tracking system using smartphone and SMS service," *2017 4th International Conference on Advances in Electrical Engineering (ICAEE)*, Dhaka,, pp. 607-612, 2017.

[10] K. V. Bakade, S. N. Parmar, S. Nainan "Wireless Communication using XBee", International Conference in Recent Trends in Information Technology and Computer Science(ICRTITCS), IJCA No. 0975 – 8887, 2012.

[11] D. Sudharsan, S. Katta" IVTrace: A Cost-Effective Vehicle Tracking System-A Prototype", International Journal of Engineering and Technology, Volume 2 No. 7, July 2012.

[12] M. A. Al-Khedher, "Hybrid GPS-GSM Localization of Automobile Tracking System", International Journal of Computer Science and Information Technology (IJCSIT), Vol 3, No 6, Dec 2011.

[13] http://www.gpsinformation.org/dale/nmea.htm.